\documentclass[12pt]{article}
\usepackage[left=2.5cm,top=2.50cm,right=2.5cm,bottom=2.50cm]{geometry}
\usepackage{mathrsfs}
\usepackage{amsmath,amssymb,latexsym,color,cancel,graphicx,bbm,colortbl}
\usepackage[english]{babel}
\usepackage[latin1]{inputenc}
\usepackage{ragged2e}
\usepackage{cite}
\usepackage{float}

\begin{document}
\date{}

\title{Quantum Dynamics and Collapse-and-Revival Phenomena in the Dunkl Anharmonic Oscillator}
\author{D. Ojeda-Guill\'en$^{a}$\footnote{{\it E-mail address:} dojedag@ipn.mx}, R. D. Mota$^{b}$ and M. Salazar-Ram\'irez$^{a}$} \maketitle

\begin{minipage}{0.9\textwidth}
\small $^{a}$ Escuela Superior de C\'omputo, Instituto Polit\'ecnico Nacional,
Av. Juan de Dios B\'atiz esq. Av. Miguel Oth\'on de Mendiz\'abal, Col. Lindavista,
Alc. Gustavo A. Madero, C.P. 07738, Ciudad de M\'exico, Mexico.\\

\small $^{b}$ Escuela Superior de Ingenier{\'i}a Mec\'anica y El\'ectrica, Unidad Culhuac\'an,
Instituto Polit\'ecnico Nacional, Av. Santa Ana No. 1000, Col. San
Francisco Culhuac\'an, Alc. Coyoac\'an, C.P. 04430, Ciudad de M\'exico, Mexico.\\

\end{minipage}

\begin{abstract}
We study the Dunkl anharmonic oscillator (Kerr medium) Hamiltonian from an algebraic approach of the $SU(1,1)$ group. In order to obtain the exact energy spectrum of this problem, we write its Hamiltonian in terms of the Dunkl creation and annihilation operators, which close the $su(1,1)$ Lie algebra. This allows us to exactly solve this Hamiltonian and obtain its parity-dependent energy spectrum. Then, we investigate the quantum dynamics of the system, particularly the collapse and revival phenomena, by using an initial state given by a superposition of even and odd Dunkl coherent states. We compute the field quadrature and the survival probability, showing that the Dunkl parameter $\mu$ modulates the fractional revivals and produces perfect state reconstructions at half-periods for specific deformation values. We analyze the quadrature variance to show that the Dunkl deformation generates interference-induced squeezed states around $t \approx \pi$. The standard Kerr medium dynamics are exactly recovered in the limit $\mu \rightarrow 0$.
\end{abstract}

\section{Introduction}

The algebraic method has been a fundamental tool to describe the non-classical properties of light in quantum optics. Systems such as coupled oscillators and the parametric amplifier have been studied using the $su(1,1)$ and $su(2)$ Lie algebras \cite{estes1968,caves1981,gerry1987}. Since Perelomov introduced the generalized coherent states for arbitrary Lie groups \cite{perelomov1972}, these methods have shown that tilting transformations can diagonalize these Hamiltonians in order to obtain their exact energy spectra and geometric phases \cite{wodkiewicz1985,gerry1989}. Also, group theory and invariant methods are essential to study the evolution of squeezed states \cite{vourdas1990}, as well as to design non-linear interferometers and quantum metrology protocols based on the $su(1,1)$ symmetry \cite{yurke1986}.

In the last decade, exact algebraic tools have been applied to analyze complex quantum systems. Exact solutions were found for the non-degenerate parametric amplifier \cite{nos2014}, the $k$-photon Jaynes-Cummings model \cite{choreno2018}, and time-dependent coupled quantum oscillators \cite{makarov2018}. Recently, there is a growing interest in applying these techniques to modern quantum optics problems. For instance, Lie algebraic treatments have been used to obtain exact analytical solutions for generalized atom-field interactions, such as non-linear optomechanical systems \cite{ramos2021}. Other applications include the study of parity-deformed Tavis-Cummings models \cite{algarni2022}, and photon sources with time-dependent Mandel $Q$ parameters \cite{jones2023}. More recently, these approaches were extended to the driven Jaynes-Cummings model \cite{bocanegra2024}. Based on these results, we were able to use this algebraic approach to compute Berry phases \cite{choreno2021}, and more recently to evaluate the Mandel parameter and geometric phases in quantum optical systems \cite{vega2024,vega2025}.

On the other hand, the study of quantum systems via algebraic deformations has progressed significantly since the work of Wigner \cite{wigner1950}. Yang introduced reflection operators to quantize space and established the deformed Heisenberg-Weyl algebras \cite{yang1951}. Later, Dunkl formulated a differential-difference operator related to reflection groups, known as the Dunkl derivative \cite{dunkl1989,dunkl2001}. Based on these foundational deformations, de Lima Rodrigues and collaborators extensively developed the operator techniques of the Wigner-Heisenberg algebra. Their work demonstrated that its generalized quadratic operators exactly close the $su(1,1)$ Lie algebra, and successfully applied this formalism to supersymmetric oscillators, canonical supercoherent states, and superconformal quantum mechanics \cite{jayaraman1990,jayaraman1999,rodrigues2003,delimarodrigues2009,carrion2010,rodrigues2013}. Within this algebraic framework, Plyuschay showed that deformed Heisenberg algebras generate hidden supersymmetries in quantum mechanical systems \cite{plyuschay1996,plyuschay1997}. Following these developments, the Dunkl formalism was applied to the harmonic oscillator and Coulomb problems in two and three dimensions \cite{genest2013,genest2014a,nos2017,ghazouani2019a}. Recently, the Dunkl formalism has been used to generalize several quantum systems. These applications include spatially varying potentials, exact oscillator models, time-evolving dynamics, higher-dimensional wave equations, statistical thermodynamics, and coherent states \cite{quesne2025,raber2025,benarous2025,hamil2025a,lut2025,hocine2025,hamil2025b,salazar2026}. These results show the capability of the Dunkl formalism to study complex parity structures and new physical phenomena.

Although Dunkl operators are increasingly used in quantum optics, most studies focus on static states or basic interactions \cite{moroz2016,hamil2022b,chung2021}. Recently, we generalized the parametric amplifier using the Dunkl derivative to tune photon bunching \cite{nos2026}. However, the effect of the Dunkl deformation on non-linear optical systems is an open problem. An important example of non-linear dynamics is the Tana\'s anharmonic oscillator (Kerr medium) \cite{tanas1984}. This model generates squeezed states and exhibits collapse and revival phenomena. These features have been observed in atom-field interactions \cite{eberly1980} and superconducting cavity circuits \cite{kirchmair2013}. Also, the anharmonic oscillator is important for the generation of Schr\"odinger cat states in quantum information processing \cite{yurkeStoler1986, mirrahimi2014}. The aim of the present work is to construct the Dunkl generalization of the Tana\'s anharmonic oscillator Hamiltonian. We replace standard ladder operators with Dunkl-deformed operators to obtain an exactly solvable model that preserves the $su(1,1)$ Lie algebra. This allows us to study how the Dunkl parameter modulates the fractional revivals and the dynamic generation of squeezed states.

This paper is organized as follows. In Section 2, we present the fundamental properties of the Dunkl $su(1,1)$ Lie algebra and its deformed operators. In Section 3, we construct the Dunkl anharmonic oscillator Hamiltonian and emphasize the exact preservation of the reflection symmetry. In Section 4, we obtain the exact energy spectrum of the Hamiltonian and analyze the collapse and revival phenomena, highlighting the fractional revivals and perfect state reconstructions at half-periods. Section 5 is dedicated to computing the temporal evolution of the field quadrature variance. Here, we explore the generation of interference-induced squeezed states via the Dunkl parameter and analyze how both the macroscopic limit and strong deformations eliminate this quantum noise reduction. Finally, our concluding remarks are given in Section 6.

\section{The Dunkl $su(1,1)$ Lie Algebra}

The Dunkl derivative is a differential-difference operator that extends the conventional derivative with a parity-dependent term. This operator leads to a deformed Heisenberg-Weyl algebra and is defined as \cite{dunkl1989,dunkl2001}
\begin{equation}
D_{\mu} = \frac{d}{dx} + \frac{\mu}{x}(1 - R),
\end{equation}
where $\mu$ is the Dunkl parameter, and $R$ is the reflection operator whose action on a function is $Rf(x) = f(-x)$. Based on this derivative, we define the Dunkl annihilation and creation operators as \cite{genest2013}
\begin{equation}
a_{\mu} = \frac{1}{\sqrt{2}}\left(x + D_{\mu}\right), \quad\quad a_{\mu}^\dagger = \frac{1}{\sqrt{2}}\left(x - D_{\mu}\right).
\end{equation}
These operators act on the Dunkl Fock states $|n\rangle$, which are the stationary states of the standard Dunkl oscillator. The action of the creation and annihilation operators on this number basis is generalized in terms of the Dunkl integers as \cite{genest2013}
\begin{subequations}
\begin{align}
a_\mu |n\rangle &= \sqrt{[n]_\mu} |n-1\rangle, \\
a_\mu^\dagger |n\rangle &= \sqrt{[n+1]_\mu} |n+1\rangle,
\end{align}
\end{subequations}
where the Dunkl integer $[n]_\mu$ is a parameter that depends on the parity of the state, defined by the relation
\begin{equation}
[n]_\mu = n + \mu\left(1 - (-1)^n\right).
\end{equation}
By successive application of the creation operator on the vacuum state $|0\rangle$, we can construct any excited state. Furthermore, the generalized Dunkl factorial is naturally defined from this action as $[n]_\mu! = [n]_\mu [n-1]_\mu \cdots [1]_\mu$, with $[0]_\mu! = 1$. The deformed operators $a_{\mu}$ and $a_{\mu}^\dagger$ satisfy the following commutation relations
\begin{equation}
[a_{\mu}, a_{\mu}^\dagger] = 1 + 2\mu R, \hspace{1.5cm} [a_{\mu}, R] = 2a_{\mu}R, \hspace{1.5cm} [a_{\mu}^\dagger, R] = 2a_{\mu}^\dagger R.
\end{equation}
Notice that the latter two commutators lead to the anti-commutation relation $\{R, a_{\mu}\} = 0$.

Also, we define the Dunkl number operator as $N_{\mu} = a_{\mu}^\dagger a_{\mu}$, which satisfies the commutation properties
\begin{equation}
[N_{\mu}, a_{\mu}] = -a_{\mu}(1 - 2\mu R), \hspace{2cm} [N_{\mu}, a_{\mu}^\dagger] = a_{\mu}^\dagger(1 + 2\mu R).
\end{equation}
In the Dunkl formalism, the standard scalar product must be modified. In order to ensure that the creation and annihilation operators $a_\mu^\dagger$ and $a_\mu$ are strictly Hermitian conjugates of each other, the inner product is defined with the weight function $|x|^{2\mu}$ as \cite{genest2013}
\begin{equation}
\langle \psi | \phi \rangle_\mu = \int_{-\infty}^{\infty} \psi^*(x) \phi(x) |x|^{2\mu} dx.
\end{equation}
The Dunkl number states $|n\rangle$ form an orthonormal basis with respect to this deformed inner product, satisfying $\langle n | m \rangle_\mu = \delta_{n,m}$. Therefore, all the expectation values and transition amplitudes computed in the following sections are implicitly evaluated under this weighted measure.

Furthermore, the Dunkl formalism allows the exact construction of coherent states. In the Glauber approach, the Dunkl coherent states $|\alpha\rangle_\mu$ are defined as the eigenstates of the annihilation operator $a_{\mu}$,
\begin{equation}
a_\mu |\alpha\rangle_\mu = \alpha |\alpha\rangle_\mu,
\end{equation}
where $\alpha$ is a complex parameter. By expanding this state in the Dunkl number basis and applying the action of the annihilation operator, the coherent states are explicitly written as \cite{ghazouani2022,sedaghatnia2023}
\begin{equation}
|\alpha\rangle_\mu = \mathcal{N}_\alpha \sum_{n=0}^\infty \frac{\alpha^n}{\sqrt{[n]_\mu!}} |n\rangle,
\end{equation}
where $\mathcal{N}_\alpha$ is the normalization constant defined in terms of the generalized Dunkl exponential function. These states minimize the generalized Heisenberg uncertainty principle and form an overcomplete basis.

Now, we introduce the quadratic operators associated with the Dunkl formalism as follows
\begin{equation}\label{Kdef}
K_+^{\mu} = \frac{1}{2}(a_{\mu}^\dagger)^2, \quad\quad K_-^{\mu} = \frac{1}{2}a_{\mu}^2, \quad\quad K_0^{\mu} = \frac{1}{4}(a_{\mu}^\dagger a_{\mu} + a_{\mu} a_{\mu}^\dagger).
\end{equation}
By substituting these dimensionless operators into the definitions of Eq. (\ref{Kdef}), we obtain their explicit differential realization as
\begin{align}
K_0^\mu &= \frac{1}{4} \left( x^2 - D_\mu^2 \right), \\
K_+^\mu &= \frac{1}{4} \left( x^2 + D_\mu^2 - \{x, D_\mu\} \right), \\
K_-^\mu &= \frac{1}{4} \left( x^2 + D_\mu^2 + \{x, D_\mu\} \right),
\end{align}
where $\{x, D_{\mu}\} = x D_{\mu} + D_{\mu} x$ is the anticommutator. Furthermore, the generators $K_{0}^{\mu}$, $K_{+}^{\mu}$, and $K_{-}^{\mu}$ can be explicitly written in terms of the standard derivative as
\begin{align}
K_{0}^{\mu} &= \frac{1}{4} \left( -\frac{d^{2}}{dx^{2}} + x^{2} - \frac{2\mu}{x}\frac{d}{dx} + \frac{\mu}{x^{2}}(1 - R) \right), \\
K_{+}^{\mu} &= \frac{1}{4} \left( \frac{d^{2}}{dx^{2}} + x^{2} + \left(\frac{2\mu}{x} - 2x\right)\frac{d}{dx} - \frac{\mu}{x^{2}}(1 - R) - (2\mu + 1) \right), \\
K_{-}^{\mu} &= \frac{1}{4} \left( \frac{d^{2}}{dx^{2}} + x^{2} + \left(\frac{2\mu}{x} + 2x\right)\frac{d}{dx} - \frac{\mu}{x^{2}}(1 - R) + (2\mu + 1) \right).
\end{align}
This explicit realization reveals that the Dunkl parameter introduces a parity-dependent centrifugal barrier term $\frac{\mu}{x^{2}}(1-R)$. This term vanishes for even parity states ($R=1$) but creates a repulsive potential for odd parity states ($R=-1$).

The set of operators $K_+^{\mu}$, $K_-^{\mu}$, and $K_0^{\mu}$ satisfy the commutation relations of the $su(1,1)$ Lie algebra \cite{nos2026}
\begin{equation}\label{Comm}
[K_0^{\mu}, K_+^{\mu}] = +K_+^{\mu}, \quad\quad [K_0^{\mu}, K_-^{\mu}] = -K_-^{\mu}, \quad\quad [K_-^{\mu}, K_+^{\mu}] = 2K_0^{\mu}.
\end{equation}
The action of the $su(1,1)$ generators on the Dunkl number states $|n\rangle$ is given by
\begin{align}
    K_{0}^{\mu}|n\rangle &= \frac{1}{2}\left([n]_{\mu}+\frac{1}{2}+\mu(-1)^{n}\right)|n\rangle, \label{K0n} \\
    K_{+}^{\mu}|n\rangle &= \frac{1}{2}\sqrt{[n+1]_{\mu}[n+2]_{\mu}}|n+2\rangle, \label{K+n} \\
    K_{-}^{\mu}|n\rangle &= \frac{1}{2}\sqrt{[n]_{\mu}[n-1]_{\mu}}|n-2\rangle. \label{K-n}
\end{align}

The Dunkl-Casimir operator commutes with all the generators of the $su(1,1)$ algebra and is defined as $C^{\mu}=(K_{0}^{\mu})^{2}-K_{0}^{\mu}-K_{+}^{\mu}K_{-}^{\mu}$. Its action on the Dunkl number states $|n\rangle$ is explicitly given by
\begin{equation}
    C^{\mu}|n\rangle = \left(\frac{\mu^{2}}{4}-\frac{\mu}{4}R-\frac{3}{16}\right)|n\rangle = k(k-1)|n\rangle.
\end{equation}
Therefore, the Bargmann indices $k_{+}$ (for the even sector $R=1$) and $k_{-}$ (for the odd sector $R=-1$) are written as
\begin{equation}
    k_{+}=\frac{1}{4}+\frac{\mu}{2}, \hspace{2cm} k_{-}=\frac{3}{4}+\frac{\mu}{2}.
\end{equation}

Notice that although the commutation relations of the operators $a_{\mu}$ and $a_{\mu}^\dagger$ depend on the reflection operator $R$, this dependence disappears in the commutators of the $su(1,1)$ quadratic operators. This is due to the anti-commutation relation $\{R, a_{\mu}\} = 0$, which ensures the exact cancellation of all $\mu$-dependent terms in the bilinear products. Therefore, the Dunkl operators $K_+^{\mu}, K_-^{\mu}, K_0^{\mu}$ preserve the structure of the $su(1,1)$ Lie algebra.

\section{The Dunkl Anharmonic Oscillator Hamiltonian}

In quantum optics, the anharmonic oscillator describes a system where the energy levels are not equally spaced due to the self-interaction of the field. The standard Hamiltonian for this system in natural units where $\hbar = 1$ and the mass of the oscillator is $m=1$ is given by \cite{tanas1984}
\begin{equation}
H = \omega a^\dagger a + \frac{\lambda}{2} (a^\dagger)^2 a^2,
\end{equation}
where $\omega$ is the field frequency and $\lambda$ is the anharmonicity or Kerr coupling constant. We construct the Dunkl generalization of this Hamiltonian by replacing the standard creation and annihilation operators by their corresponding generalization with the Dunkl derivative
\begin{equation}\label{Hu}
H_{\mu} = \omega a_{\mu}^\dagger a_{\mu} + \frac{\lambda}{2} (a_{\mu}^\dagger)^2 a_{\mu}^2.
\end{equation}
This Hamiltonian models non-linear optical processes in systems with reflection symmetry.

In order to write the Hamiltonian in terms of the $su(1,1)$ generators, we use the following equalities
\begin{equation}
a_{\mu}^\dagger a_{\mu} = 2K_0^{\mu} - \frac{1}{2} - \mu R, \quad\quad (a_{\mu}^\dagger)^2 a_{\mu}^2 = 4K_{+}^{\mu} K_{-}^{\mu}.
\end{equation}
By substituting these relations into Eq. (\ref{Hu}), we obtain the Hamiltonian in terms of the $su(1,1)$ generators as
\begin{equation}\label{Ham}
H_{\mu} = \omega \left( 2K_0^{\mu} - \frac{1}{2} - \mu R \right) + 2\lambda K_{+}^{\mu} K_{-}^{\mu}.
\end{equation}
Unlike standard quantum systems, the constant term $-\omega(1/2 + \mu R)$ cannot be neglected, since it introduces a parity-dependent phase shift that fundamentally modifies the time evolution of the system.

On the other hand, the Dunkl number operator $N_{\mu} = a_{\mu}^\dagger a_{\mu}$ is a constant of motion, as it satisfies the commutation relation
\begin{equation}
[H_{\mu}, N_{\mu}] = 0.
\end{equation}
Since the number of excitations is conserved, the photon number distribution remains stationary throughout the evolution. Therefore, the Mandel $Q$ parameter is constant in time. For this reason, our study focuses on the field quadratures and the squeezing properties, which are highly sensitive to the non-linear phase accumulation.

Finally, the Hamiltonian commutes with the reflection operator $R$ as
\begin{equation}
[H_{\mu}, R] = \omega [N_\mu, R] + 2\lambda [K_{+}^\mu K_{-}^\mu, R] = 0.
\end{equation}
Therefore, the reflection symmetry is preserved, and the eigenfunctions of the system are split into two independent parity sectors: the even sector ($R=1$) and the odd sector ($R=-1$). This separation provides the basis for our exact spectral and dynamical analysis.

\section{Exact energy spectrum and quantum dynamics}

In this section, we obtain the exact energy spectrum of the Dunkl anharmonic oscillator and investigate its quantum dynamics. The $su(1,1)$ algebraic structure allows us to exactly solve the Hamiltonian and describe the collapse and revival phenomena.

\subsection{Energy spectrum and parity splitting}

In order to obtain the energy spectrum, we write the product $K_+^\mu K_-^\mu$ in terms of the Casimir operator $C^\mu$ and the generator $K_0^\mu$. From the definition $C^\mu = (K_0^\mu)^2 - K_0^\mu - K_+^\mu K_-^\mu$, we substitute this expression into the Hamiltonian Eq. (\ref{Ham}) to obtain
\begin{equation}
H_{\mu} = 2\lambda (K_{0}^{\mu})^{2} + 2(\omega - \lambda)K_{0}^{\mu} - 2\lambda C^{\mu} - \omega \left( \mu R + \frac{1}{2} \right).
\end{equation}
By applying this Hamiltonian to the Dunkl number states $|n\rangle$ and using the eigenvalues of $K_0^\mu$ and $C^\mu$, we obtain the exact energy spectrum
\begin{equation}
E_n^\mu = 2\lambda \left[ \frac{1}{2}\left([n]_\mu + \frac{1}{2} + \mu(-1)^n\right) \right]^2 + (\omega - \lambda)\left([n]_\mu + \frac{1}{2} + \mu(-1)^n\right) - 2\lambda k(k-1) - \omega \left( \mu (-1)^n + \frac{1}{2} \right).
\end{equation}
As stated at the end of Section 3, the preservation of the reflection symmetry splits the energy spectrum into two independent parity sectors. For the even sector ($n=2m$), the energy levels are given by
\begin{equation}
E_{2m}^\mu = 2\lambda m^2 + m(2\omega + 2\lambda\mu - \lambda),
\end{equation}
and for the odd sector ($n=2m+1$), we obtain
\begin{equation}
E_{2m+1}^\mu = 2\lambda m^2 + m(2\omega + 2\lambda\mu + \lambda) + \omega(1+2\mu).
\end{equation}
Notice that the Dunkl parameter $\mu$ introduces a linear shift in $m$ and a global energy displacement that depends on the parity of the state.

Now, we verify that our generalized energy spectrum reduces to the standard Kerr medium in the limit $\mu \rightarrow 0$. In this limit, the parity sectors recombine into the standard integer photon number $n$. For the even sector ($n=2m \implies m=n/2$), the energy spectrum becomes
\begin{equation}
E_n^{(e)} = 2\lambda \left(\frac{n}{2}\right)^2 + \frac{n}{2}(2\omega - \lambda) = \omega n + \frac{\lambda}{2}n(n-1).
\end{equation}
For the odd sector ($n=2m+1 \implies m=(n-1)/2$), the energy reduces to
\begin{equation}
E_n^{(o)} = 2\lambda \left(\frac{n-1}{2}\right)^2 + \frac{n-1}{2}(2\omega + \lambda) + \omega = \omega n + \frac{\lambda}{2}n(n-1).
\end{equation}
Notice that both sectors converge to the same expression, which matches the standard eigenvalues of the anharmonic oscillator \cite{tanas1984,yurkeStoler1986}. Therefore, our algebraic approach is consistent in both parity sectors.

\subsection{Collapse and revival phenomena}

In order to study the quantum dynamics, we select an initial state with a classical-like behavior. We use the Dunkl coherent state $|\alpha\rangle_\mu$ introduced in Section 2, which satisfies the eigenvalue equation $a_\mu |\alpha\rangle_\mu = \alpha |\alpha\rangle_\mu$ \cite{ghazouani2022,sedaghatnia2023}. Since parity is a constant of motion, the time evolution of the even and odd sectors is decoupled. Therefore, we write the initial state as a superposition of the even and odd parity sectors of the Dunkl coherent state as
\begin{equation}
|\psi(0)\rangle = \mathcal{N} \left( \sum_{m=0}^\infty \frac{\alpha^{2m}}{\sqrt{[2m]_\mu!}} |2m\rangle + \sum_{m=0}^\infty \frac{\alpha^{2m+1}}{\sqrt{[2m+1]_\mu!}} |2m+1\rangle \right),
\end{equation}
where $\mathcal{N}$ is the new normalization constant for the superposed state. It is worth noting that in the standard Kerr medium ($\mu=0$), the initial coherent state is typically written as a single continuous sum over all photon numbers $n$, since the energy spectrum does not depend on the parity of the states. However, in the Dunkl formalism, the exact preservation of the reflection symmetry forces the energy spectrum to split into two independent sets of eigenvalues for the even and odd sectors. Therefore, to correctly compute the temporal evolution, the initial coherent state must be explicitly constructed as a superposition of these two parity branches. Furthermore, this explicit superposition is physically necessary to observe the oscillatory dynamics of the field, since a state of definite parity yields an expectation value of the field quadrature equal to zero at all times.

By applying the time evolution operator $U(t) = \exp(-iH_\mu t)$ to the initial state, we determine the dynamics of the system. Since the Dunkl number states are exact eigenstates of the Hamiltonian, the action of the evolution operator yields a phase accumulation given by $U(t)|n\rangle = \exp(-iE_n^\mu t)|n\rangle$. Therefore, we obtain the state at any time $t$ as
\begin{equation}
|\psi(t)\rangle = \mathcal{N} \left( \sum_{m=0}^\infty \frac{\alpha^{2m}}{\sqrt{[2m]_\mu!}} e^{-i E_{2m}^\mu t} |2m\rangle + \sum_{m=0}^\infty \frac{\alpha^{2m+1}}{\sqrt{[2m+1]_\mu!}} e^{-i E_{2m+1}^\mu t} |2m+1\rangle \right).
\end{equation}
The non-linear term $m^2$ in the energy spectra $E_{2m}^\mu$ and $E_{2m+1}^\mu$ causes the dispersion of the wave packet, which leads to the collapse of the signal and its periodic revivals.

In order to study this temporal evolution, we compute the expectation value of the field quadrature, which is defined as
\begin{equation}
X_\mu = \frac{1}{\sqrt{2}}(a_\mu + a_\mu^\dagger).
\end{equation}
First, we evaluate the expectation value of the annihilation operator. Because the action of this operator lowers the quantum number by one, it strictly exchanges the parity of the states. By applying $a_\mu$ to the time-evolved state $|\psi(t)\rangle$ and computing the inner product with $\langle \psi(t) |$, the orthogonality condition eliminates all terms except the cross-terms between the even and odd sectors. Thus, we obtain
\begin{equation}
\langle \psi(t) | a_\mu | \psi(t) \rangle = \mathcal{N}^2 \sum_{m=0}^\infty \left[ \frac{\alpha^{4m+1}}{[2m]_\mu!} e^{-i(E_{2m+1}^\mu - E_{2m}^\mu) t} + \frac{\alpha^{4m+3}}{[2m+1]_\mu!} e^{-i(E_{2m+2}^\mu - E_{2m+1}^\mu) t} \right].
\end{equation}
Since the creation operator $a_\mu^\dagger$ is the Hermitian adjoint of $a_\mu$, its expectation value is simply the complex conjugate $\langle a_\mu^\dagger \rangle = \langle a_\mu \rangle^*$. Therefore, the expectation value of the quadrature can be written as $\langle X_\mu(t) \rangle = \sqrt{2} \text{Re}[\langle a_\mu(t) \rangle]$. By assuming the coherent parameter $\alpha$ to be real, the real part of the temporal exponentials yields the cosine functions. This expectation value is explicitly given by
\begin{equation}\label{X_expectation}
\langle X_\mu(t) \rangle = \sqrt{2} \mathcal{N}^2 \sum_{m=0}^\infty \left[ \frac{\alpha^{4m+1}}{[2m]_\mu!} \cos\left( (E_{2m+1}^\mu - E_{2m}^\mu) t \right) + \frac{\alpha^{4m+3}}{[2m+1]_\mu!} \cos\left( (E_{2m+2}^\mu - E_{2m+1}^\mu) t \right) \right].
\end{equation}

\begin{figure}[tbp]
    \centering
    \includegraphics[width=1\textwidth]{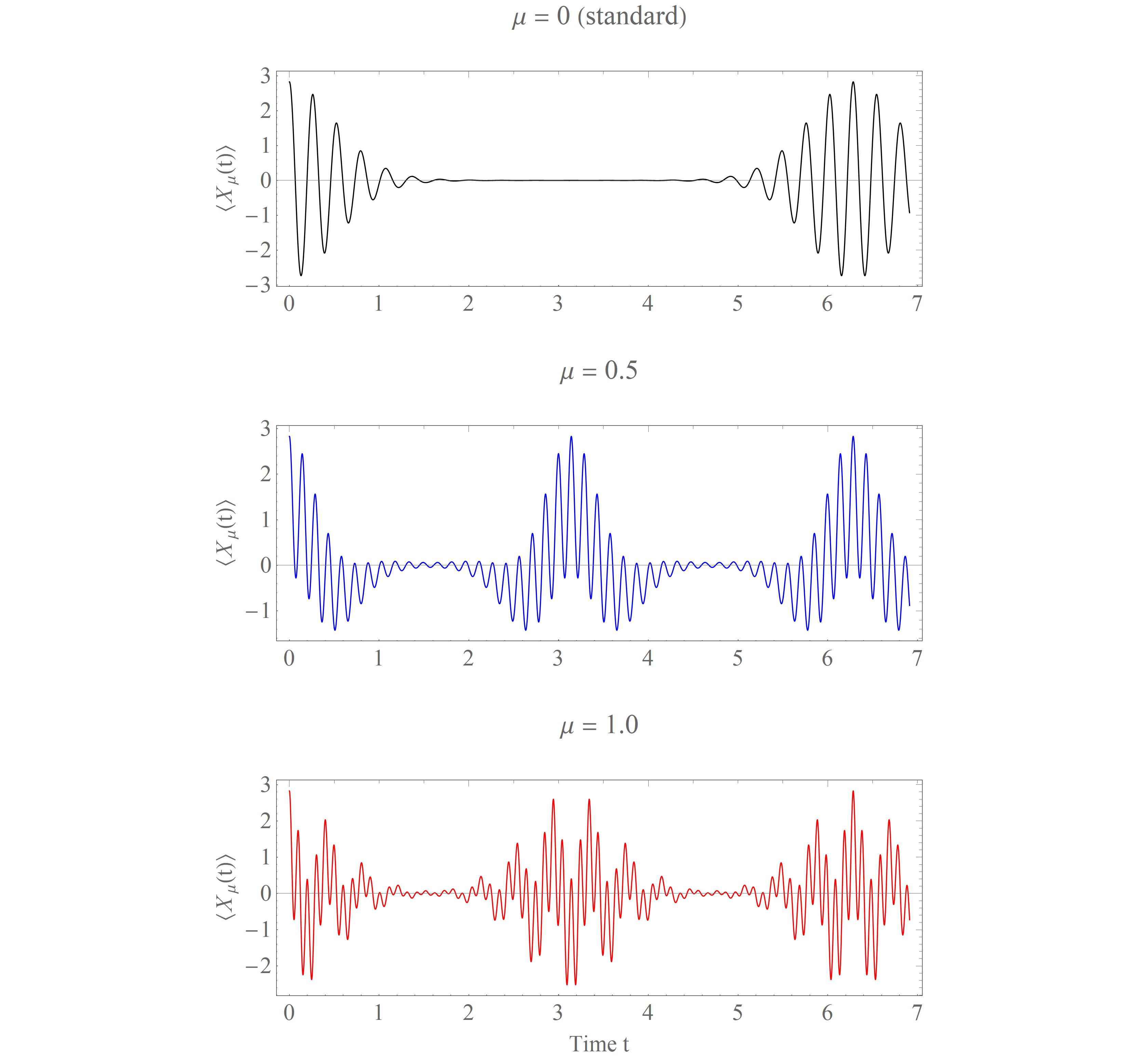}
    \caption{Evolution of the quadrature $\langle X_{\mu}(t)\rangle$ over a full cycle. The Dunkl parameter $\mu$ produces a half-cycle revival at $t\approx\pi$, while a complete resynchronization of all wave packets occurs at $t\approx2\pi$. For this numerical simulation, the anharmonicity parameter is set to $\lambda=1$, the free oscillator frequency to $\omega=20$, and the initial coherent state amplitude is $\alpha=2$.}
    \label{fig:quadrature}
\end{figure}

In Fig. \ref{fig:quadrature} we show the temporal evolution of the field quadrature. We observe that the Dunkl parameter significantly modifies the dynamics at half the cycle. Specifically, at $t = \pi$, the deformed systems ($\mu=0.5$ and $\mu=1.0$) exhibit distinct fractional revivals, whereas the standard case ($\mu=0$) remains completely collapsed. However, a complete resynchronization of the wave packets occurs at $t = 2\pi$ for all values of $\mu$. This confirms that while the Dunkl deformation produces a half-cycle revival, it strictly preserves the global periodicity of the Kerr medium.

Now, we quantify the revival dynamics through the survival probability (fidelity), defined as $F(t) = |\langle \psi(0) | \psi(t) \rangle|^2$. In order to evaluate this expression, we first write the dual of the initial state (the bra vector) as
\begin{equation}
\langle \psi(0) | = \mathcal{N} \left( \sum_{m=0}^\infty \frac{(\alpha^*)^{2m}}{\sqrt{[2m]_\mu!}} \langle 2m| + \sum_{m=0}^\infty \frac{(\alpha^*)^{2m+1}}{\sqrt{[2m+1]_\mu!}} \langle 2m+1| \right).
\end{equation}
By computing the inner product with the time-evolved state $|\psi(t)\rangle$, we use the orthogonality condition of the Dunkl number states, $\langle n | m \rangle = \delta_{n,m}$. This condition ensures that all cross-terms between different states vanish. Therefore, the inner product reduces strictly to the multiplication of the corresponding coefficients in the even and odd sectors. Since $\alpha^* \alpha = |\alpha|^2$, we obtain the exact overlap amplitude as
\begin{equation}
\langle\psi(0)|\psi(t)\rangle = \mathcal{N}^2 \left( \sum_{m=0}^\infty \frac{|\alpha|^{4m}}{[2m]_\mu!} e^{-i E_{2m}^\mu t} + \sum_{m=0}^\infty \frac{|\alpha|^{4m+2}}{[2m+1]_\mu!} e^{-i E_{2m+1}^\mu t} \right).
\end{equation}
Thus, the fidelity $F(t)$ is simply the squared modulus of this complex amplitude.

\begin{figure}[tbp]
    \centering
    \includegraphics[width=1\textwidth]{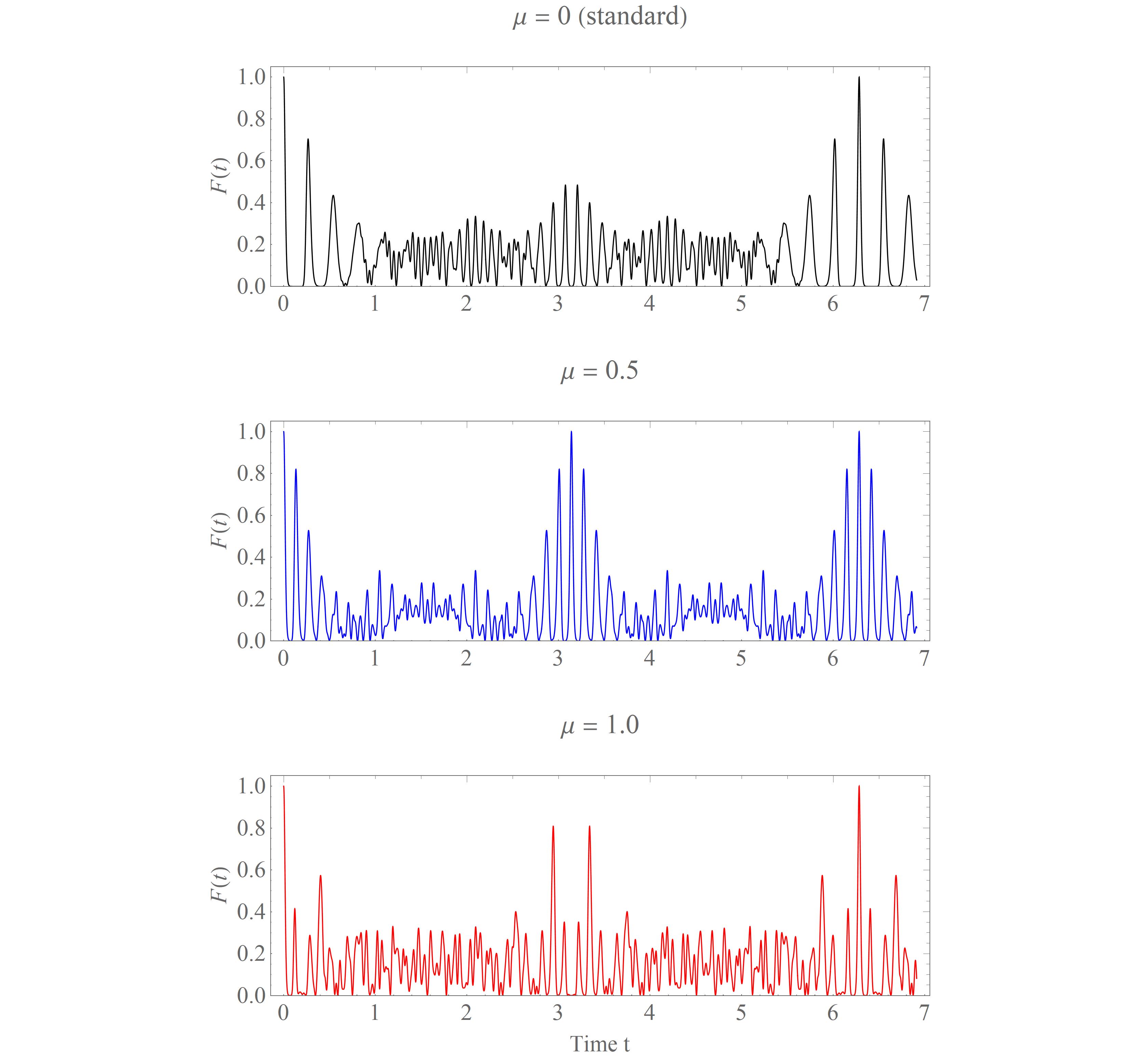}
    \caption{Survival probability $F(t)$ for different values of the Dunkl parameter: $\mu=0$ (standard), $\mu=0.5$ and $\mu=1.0$. The vertical alignment at $t\approx6.28$ confirms the universal revival period $2\pi/\lambda$ (evaluated here with the free oscillator frequency $\omega=20$, the anharmonicity parameter $\lambda=1$, and the initial amplitude $\alpha=2$). For $\mu=0.5$, a perfect additional revival occurs at $t=\pi/\lambda$.}
    \label{fig:fidelity}
\end{figure}

In Fig. \ref{fig:fidelity} we show the numerical results for the survival probability $F(t)$. Notice that the fundamental revival period $t_R = 2\pi/\lambda$ remains invariant for all $\mu$, but the Dunkl deformation modulates the fractional revivals. For semi-integer values such as $\mu=0.5$, the Dunkl deformation produces a new perfect revival at $t = \pi/\lambda$. Therefore, the Dunkl parameter $\mu$ tunes the revival phase, which could be applied to generate new displaced cat states for quantum information processing.

\section{Quantum statistics and squeezing dynamics}

Since the Dunkl number operator is a constant of motion ($[H_\mu, N_\mu]=0$), the photon number distribution remains constant in time. Therefore, the Mandel $Q$ parameter is strictly constant. In order to study the dynamical deformation of the quantum noise, we compute the variance of the field quadrature.

The squeezing of the quantum field is characterized by the variance of the Dunkl quadrature $X_\mu$. Notice that a state exhibits squeezing if its variance falls below the standard quantum limit, i.e., $(\Delta X_\mu)^2 < 0.5$. The variance is defined as
\begin{equation}
(\Delta X_\mu(t))^2 = \langle X_\mu^2(t) \rangle - \langle X_\mu(t) \rangle^2.
\end{equation}
The expectation value $\langle X_\mu(t) \rangle$ was computed in Eq. (\ref{X_expectation}). In order to evaluate $\langle X_\mu^2(t) \rangle$, we write $X_\mu^2$ in terms of the $su(1,1)$ generators as
\begin{equation}
X_\mu^2 = \frac{1}{2}\left(a_\mu^2 + (a_\mu^\dagger)^2 + a_\mu a_\mu^\dagger + a_\mu^\dagger a_\mu\right) = K_-^\mu + K_+^\mu + 2K_0^\mu.
\end{equation}
By computing the expectation value, we obtain
\begin{equation}\label{X2exp}
\langle X_\mu^2(t) \rangle = 2\text{Re}[\langle K_-^\mu(t) \rangle] + 2\langle K_0^\mu \rangle.
\end{equation}

In order to evaluate $\langle K_-^\mu(t) \rangle$, we first analyze the action of the operator $K_-^\mu = \frac{1}{2}a_\mu^2$ on the Dunkl number states. According to Eq. (\ref{K-n}), this operator lowers the generalized quantum number by two and strictly preserves the parity of the states. By applying this operator to the time-evolved state $|\psi(t)\rangle$, we obtain
\begin{equation}
K_-^\mu |\psi(t)\rangle = \frac{\mathcal{N}}{2} \left( \sum_{m=1}^\infty \frac{\alpha^{2m}}{\sqrt{[2m-2]_\mu!}} e^{-i E_{2m}^\mu t} |2m-2\rangle + \sum_{m=1}^\infty \frac{\alpha^{2m+1}}{\sqrt{[2m-1]_\mu!}} e^{-i E_{2m+1}^\mu t} |2m-1\rangle \right).
\end{equation}
Notice that the sum index starts at $m=1$ since the terms for $m=0$ are annihilated by the operator. Now, we compute the inner product with the dual state $\langle \psi(t) |$. By using the orthogonality condition of the Dunkl number states, $\langle n | m \rangle = \delta_{n,m}$, the inner product does not generate interference between parities. The non-vanishing terms strictly couple the states within the same parity sector. Since $\alpha^{2m}(\alpha^*)^{2m-2} = \alpha^2 |\alpha|^{4m-4}$, we obtain the exact analytical expression
\begin{equation}
\langle K_-^\mu(t) \rangle = \frac{\mathcal{N}^2 \alpha^2}{2} \left( \sum_{m=0}^\infty \frac{|\alpha|^{4m}}{[2m]_\mu!} e^{-i \Delta E_{e} t} + \sum_{m=0}^\infty \frac{|\alpha|^{4m+2}}{[2m+1]_\mu!} e^{-i \Delta E_{o} t} \right),
\end{equation}
where we redefined the sum index $m \rightarrow m+1$, and the phase evolution is determined by the energy differences between next-nearest neighbor states
\begin{align}
\Delta E_{e} &= E_{2m+2}^\mu - E_{2m}^\mu = 2\omega + \lambda(4m + 2\mu + 1), \\
\Delta E_{o} &= E_{2m+3}^\mu - E_{2m+1}^\mu = 2\omega + \lambda(4m + 2\mu + 3).
\end{align}

By substituting these expressions into Eq. (\ref{X2exp}), we compute the temporal evolution of the squared quadrature. In order to obtain this explicit result, we take the real part of $\langle K_-^\mu(t) \rangle$. By assuming the coherent parameter $\alpha$ to be real, the real part of the temporal exponentials yields the cosine functions. Thus, we obtain
\begin{equation}
\langle X_\mu^2(t) \rangle = \mathcal{N}^2 \alpha^2 \left( \sum_{m=0}^\infty \frac{|\alpha|^{4m}}{[2m]_\mu!} \cos(\Delta E_{e} t) + \sum_{m=0}^\infty \frac{|\alpha|^{4m+2}}{[2m+1]_\mu!} \cos(\Delta E_{o} t) \right) + 2\langle K_0^\mu \rangle.
\end{equation}

Therefore, by substituting this result into the definition of the variance, we obtain the exact analytical expression for the temporal evolution of the quantum noise as
\begin{equation}\label{VarX_explicit}
(\Delta X_\mu(t))^2 = \mathcal{N}^2 \alpha^2 \left( \sum_{m=0}^\infty \frac{|\alpha|^{4m}}{[2m]_\mu!} \cos(\Delta E_{e} t) + \sum_{m=0}^\infty \frac{|\alpha|^{4m+2}}{[2m+1]_\mu!} \cos(\Delta E_{o} t) \right) + 2\langle K_0^\mu \rangle - \langle X_\mu(t) \rangle^2.
\end{equation}
Here, the expectation value of the field quadrature $\langle X_\mu(t) \rangle$ is explicitly given by Eq. (\ref{X_expectation}). Furthermore, since $[H_\mu, K_0^\mu]=0$, the term $2\langle K_0^\mu \rangle$ is a constant of motion evaluated with respect to the initial state $|\psi(0)\rangle$. By using the eigenvalue relation of Eq. (\ref{K0n}), this constant is explicitly written as
\begin{equation}
2\langle K_0^\mu \rangle = \mathcal{N}^2 \left( \sum_{m=0}^\infty \frac{|\alpha|^{4m}}{[2m]_\mu!} \left( 2m + \mu + \frac{1}{2} \right) + \sum_{m=0}^\infty \frac{|\alpha|^{4m+2}}{[2m+1]_\mu!} \left( 2m + \mu + \frac{3}{2} \right) \right).
\end{equation}

\begin{figure}[tbp]
    \centering
    \includegraphics[width=1\textwidth]{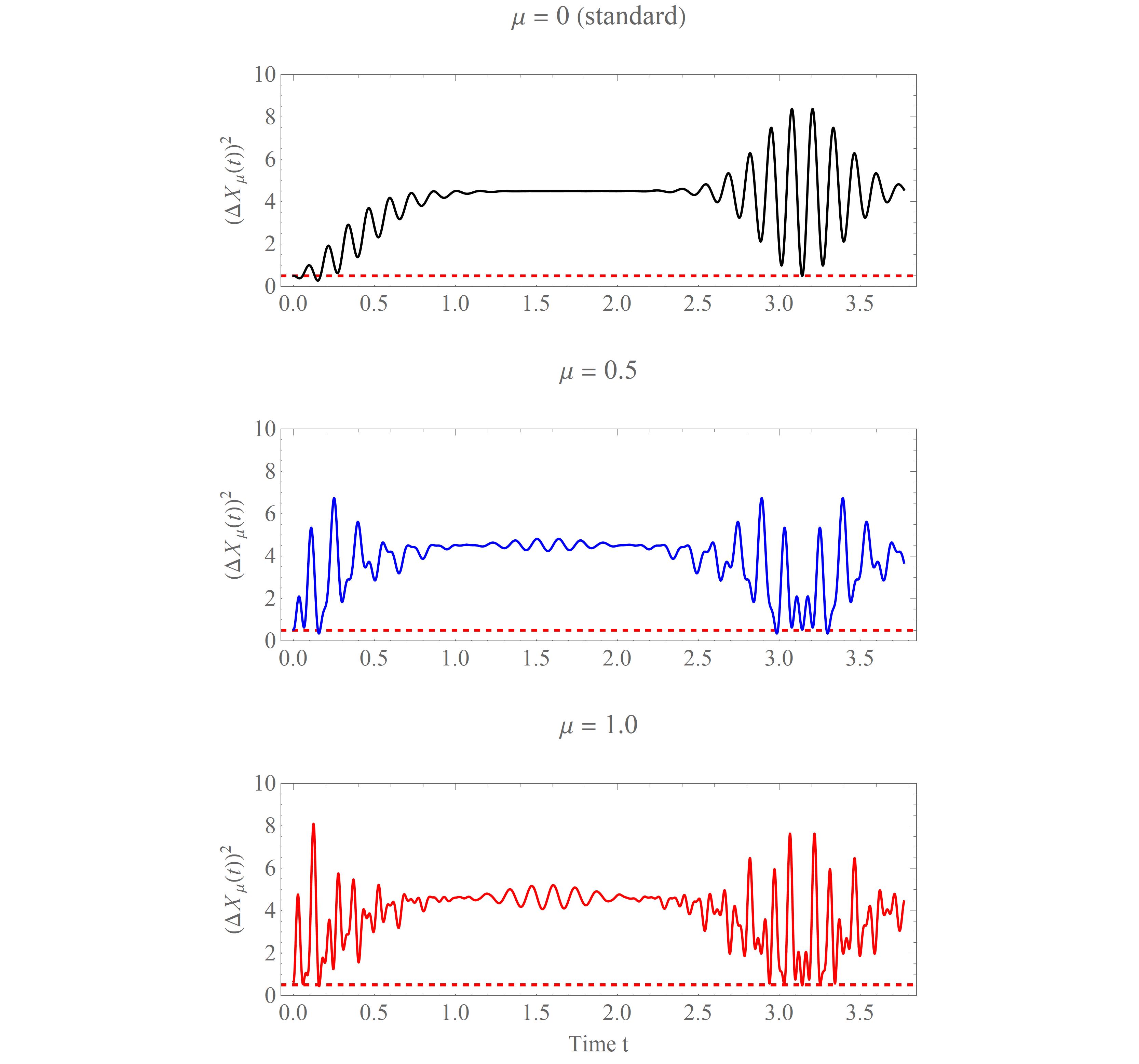}
    \caption{Temporal evolution of the quadrature variance $(\Delta X_{\mu}(t))^{2}$ for $\mu=0$ (top), $\mu=0.5$ (middle), and $\mu=1.0$ (bottom), setting the free oscillator frequency to $\omega=20$, the Kerr parameter to $\lambda=1$, and the coherent amplitude to $\alpha=2$. The red dashed horizontal line marks the Standard Quantum Limit (SQL) at $0.5$. Values below this line indicate the generation of quadrature squeezed light.}
    \label{fig:squeezing}
\end{figure}

As the state evolves, the non-linear interactions redistribute the phase space uncertainty, generating squeezed states. This process is subsequently followed by the collapse of the signal, during which the variance enters a quasi-stationary regime. This stable behavior physically corresponds to the uniform spreading of the wave packet along a ring in phase space, where the variance stabilizes at the constant value $2\langle K_0^\mu \rangle \approx |\alpha|^2 + 0.5$.

In Fig. \ref{fig:squeezing} we show the dynamics of the quadrature squeezing for different values of the Dunkl parameter. In the standard case ($\mu=0$), the quantum noise drops below the Standard Quantum Limit (SQL) only at the beginning of the evolution. However, for $\mu=0.5$, the system exhibits initial squeezing followed by two distinct squeezing dips around $t = \pi$ (one just before and one just after this time). For $\mu=1.0$, while the initial squeezing is preserved, the noise reduction around $t = \pi$ becomes irregular and essentially the squeezing disappears. Finally, analyzing the variances for different values of $\alpha$ and $\mu$ reveals that increasing either parameter eliminates both the initial and subsequent squeezing.

Finally, in the limit $\mu \rightarrow 0$, the generalized Dunkl numbers reduce to standard integers $[n]_\mu \rightarrow n$, and the Dunkl factorials become standard factorials $[n]_\mu! \rightarrow n!$. Therefore, the parity-dependent energy differences $\Delta E_{e}$ and $\Delta E_{o}$ become equal, and the expression for $\langle K_-^\mu(t) \rangle$ recovers the standard time evolution of the quantum noise. Thus, the dynamically squeezed Dunkl variance $(\Delta X_\mu(t))^2$ exactly reproduces the statistical properties of the standard Kerr medium.

\section{Concluding remarks}

In this paper, we studied the Dunkl anharmonic oscillator Hamiltonian from an algebraic approach of the $su(1,1)$ Lie algebra. In order to solve the spectral problem exactly, we wrote the Hamiltonian in terms of the Dunkl creation and annihilation operators. We obtained the exact energy spectrum and showed that the Dunkl parameter $\mu$ introduces a parity-dependent shift in the energy levels. Since the reflection symmetry is strictly preserved, the spectrum is split into two independent even and odd sectors.

Then, we studied the quantum dynamics of the system by using an initial state given by a superposition of even and odd Dunkl coherent states. The field quadrature and the survival probability were calculated to describe the collapse and revival phenomena. We showed that the fundamental revival period remains invariant for all values of $\mu$, but the Dunkl deformation modulates the fractional revivals. Specifically, for semi-integer values such as $\mu=0.5$, the Dunkl parameter generates a perfect state reconstruction at exactly half the fundamental period. Thus, the Dunkl deformation produces a half-cycle revival without destroying the global revival periodicity of the system.

Finally, we studied the quantum statistics of the evolved state. Since the Dunkl number operator is a constant of motion, the photon distribution remains constant in time. However, by computing the temporal evolution of the quadrature variance, we showed that the Dunkl deformation fundamentally modifies the quantum noise profile. While the standard Kerr medium exhibits squeezing only at the beginning of the evolution, the Dunkl deformed system generates discrete, interference-induced squeezed states around $t \approx \pi$. Furthermore, we explained that when considering the macroscopic limit (large amplitude $\alpha$) and strong deformations (large $\mu$), the numerical simulations show that this quantum noise reduction is eliminated.

If we consider the limit $\mu \rightarrow 0$, the generalized Dunkl numbers reduce to standard integers. Thus, our exact analytical results recover the energy spectrum and exactly reproduce the well-known collapse and revival statistical properties of the standard Kerr medium \cite{tanas1984,milburn1986}.

It is worth noting that the emergence of the fractional revivals in the survival probability has a profound physical significance. As demonstrated by Yurke and Stoler for the standard Kerr medium \cite{yurke1986}, the appearance of fractional peaks indicates that the initial wave packet has evolved into a macroscopic quantum superposition of distinguishable states, commonly referred to as Schr\"odinger cat states. In the context of our model, the fractional revivals observed at $t = \pi/\lambda$ and other rational fractions of the fundamental period could imply the generation of Dunkl-deformed Schr\"odinger cat states. A complete phase-space analysis of these states via the Wigner function would reveal them as a macroscopic superposition of Dunkl coherent states exhibiting distinct quantum interference fringes. A detailed exploration of this new phase-space structure of the Dunkl anharmonic oscillator will be reported in a forthcoming paper.

\section*{Acknowledgments}

This work was partially supported by SNII-M\'exico, COFAA-IPN, EDI-IPN, and CGPI-IPN Project Number 20251355.\\

\section*{Disclosures}

The authors declare no conflicts of interest.

\section*{Data Availability}

No data were generated or analyzed in the presented research.

\end{document}